\documentclass[12pt]{article}
\usepackage{graphicx}

\begin{document}
\newpage

\begin{flushright}
SLAC-PUB-9197\\
April 2002
\end{flushright}

\bigskip\bigskip

\begin{center}
{{\bf\Large Physics Opportunities at a \\[1ex]
Photon--Photon Collider\footnote{Work by the Department of Energy,
contract DE--AC03--76SF00515.}}}

\vfill
{\it Stanley J. Brodsky \\
Stanford Linear Accelerator Center \\
 2575 Sand Hill Road, Menlo
Park, CA 94025 \\
 e-mail: sjbth@slac.stanford.edu}

\end{center}

\bigskip
\vfill

\begin{center}
Invited Talk presented at the \\
4th International Workshop On Electron-Electron Interactions \\
at TeV Energies (E- E- 01) \\
University of California, Santa Cruz, California \\
7--9 December 2001
\end{center}

\vfill

\newpage
\

\renewcommand{\bar}[1]{\overline{#1}}

\begin{center}
Abstract

\end{center}

 The advent of back-scattered laser beams for
$e^\pm e^-$ colliders will allow detailed studies of a large array
of high energy $\gamma \gamma$ and $\gamma e$ collision processes
with polarized beams. These include tests of electroweak theory in
photon-photon annihilation such as $\gamma \gamma \to W^+ W^-$,
$\gamma \gamma \to $ Higgs bosons, and higher-order loop
processes, such as $\gamma \gamma \to \gamma \gamma, Z \gamma, H^0
Z^0$ and $Z Z.$ Methods for measuring the anomalous magnetic and
quadrupole moments of the $W$ and $Z$ gauge bosons to high
precision in polarized electron-photon and photon-photon
collisions are discussed. Since each photon can be resolved into a
$W^+ W^-$ pair, high energy photon-photon collisions can also
provide a remarkably background-free laboratory for studying $W W$
collisions and annihilation. I also review high energy $\gamma
\gamma$ and $e \gamma$ tests of quantum chromodynamics, including
the production of two gluon jets in photon-photon collisions,
deeply virtual Compton scattering on a photon target, and
leading-twist single-spin asymmetries for a photon polarized
normal to a production plane. Exclusive hadron production
processes in photon-photon collisions provide important tests of
QCD at the amplitude level, particularly as measures of hadron
distribution amplitudes which are also important for the analysis
of exclusive semi-leptonic and two-body hadronic $B$-decays.
\bigskip

\section{Introduction}

One of the most important areas of investigation at the next
electron-positron linear collider will be the study of
photon-photon collisions.  Since photons couple directly to all
fundamental fields carrying the electromagnetic current---%
leptons, quarks, $W's,$ supersymmetric particles, etc.---%
high energy $\gamma \gamma $ collisions will provide a
comprehensive laboratory for exploring virtually every aspect of
the Standard Model and its extensions~\cite{Brodsky:1994nf}.

When a polarized laser beam Compton-scatters on a polarized
electron beam, each electron is effectively converted into a
polarized photon with a high fraction of its energy. The effective
luminosity and energy of photon-photon collisions from
back-scattered laser beams~\cite{Ginzburg:1981ik,Telnov:1998vs} is
expected to be comparable to that of the primary electron-positron
collisions.  As reported by Asner and Early~\cite{Asner:2001vu},
the technology for the required high-powered lasers is well along
in development.  There are proposals to test this technology using
elements of the SLAC Linear Collider.  The high energy luminosity,
and polarization of the back-scattered laser beams collisions has
the potential to make photon-photon collisions a key component of
the physics program of the next linear
collider~\cite{Velasco:2002vg}. Polarized electron-photon
collisions are another important byproduct of this program.

Photon-photon collisions can be classified as follows: (A) The
photons can annihilate into a charged pair such as $\gamma \gamma
\to W^+ W^-, q \bar q$, lepton pairs or charged Higgs; (B) the
photons can produce neutral pairs via loop diagrams such as
$\gamma \gamma \to Z^0 Z^0, \gamma Z^0$  and $\gamma \gamma \to g
g$ ; or (C) the photons can each couple to separate charged pairs
which scatter by a gauge particle exchange: $\gamma \gamma \to q_1
\bar q_1 q_2 \bar q_2;$ (D) the photons can fuse to produce a
single even $C$ resonance such as a neutral Higgs, an $\eta_b,$ or
$\chi_b$ higher orbital state. Exclusive hadronic final states
such as meson or baryon pairs can be formed.  In each case, a
state of even charge conjugation $C$ is produced in a general
partial wave.

\begin{figure}[htbp]
\begin{center}
\includegraphics[height=19cm,width=12cm]{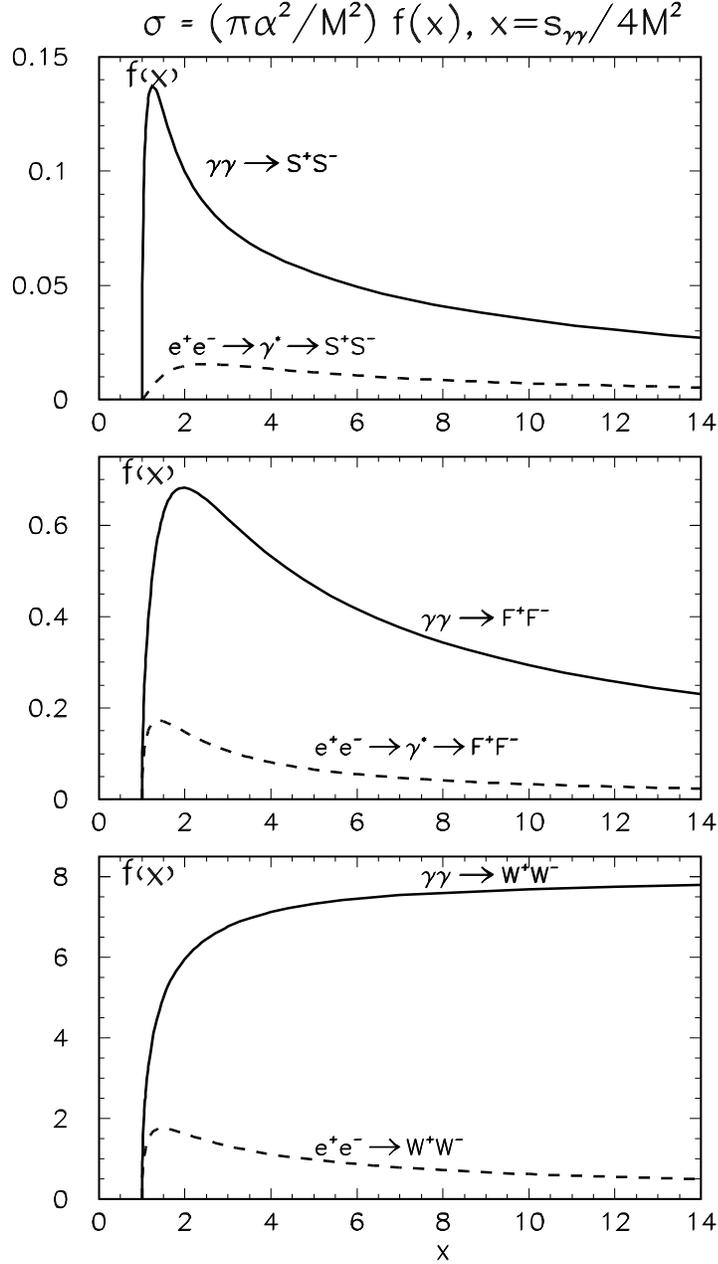}
\end{center}
\caption[*]{Comparison between cross sections for charged pair
production in unpolarized $e^+ e^-$ and $\gamma \gamma$
collisions. S (scalars), F (fermions), W ($W$ bosons); $\sqrt{s}$
is the invariant mass (c.m.s. energy of colliding beams). The
contribution of the $Z^0$ boson to the production of S and F in
$e^+ e^-$ collisions was not included. From Boos {\em et
al.}~\cite{Boos:2000ki}.\label{charged}}
\end{figure}

A unique advantage of a photon-photon collider is its potential to
produce and determine the properties of fundamental $C=+$
resonances such as the Higgs boson~\cite{Asner:2001ia}.  One can
also use the transverse polarization of the colliding photons to
distinguish the parity of the resonance:  the coupling for a
scalar resonance is $\epsilon_1\cdot\epsilon_2$ versus
$\epsilon_1\times k_1\cdot\epsilon_2$ for the pseudoscalar.  More
generally, one can use polarized photon-photon scattering to study
CP violation in the fundamental Higgs to two-photon
couplings~\cite{Grzadkowski:1992sa,Cheung:bn}.  In the case of
electron-photon collisions, one can use the transverse momentum
fall-off of the recoil electron in $ e \gamma\rightarrow e H^0$ to
measure the fall-off of the $\gamma \to $ Higgs transition form
factor and thus check the mass scale of the internal massive quark
and $W$ loops coupling to the Higgs~\cite{tang}. The cross
sections for pairs of scalars, fermions or vectors particles are
all significantly larger (by about one order of magnitude) in
$\gamma \gamma$ collisions than in $e^+e^-$ collisions, as
demonstrated in Fig~\ref{charged}.

Unlike the $e^+ e^-$ annihilation cross section, which falls at
least as fast as $1/s,$ many of the $\gamma \gamma$ cross sections
increase with energy. The energy dependence of a cross section
follows from the spin of the exchanged quanta.  Using Regge
analysis, a two-body cross section ${d\sigma\over dt} \propto s^{2
\alpha_R(t)-2}\beta(t)$ at fixed $t$ where $\alpha_R$ is the spin
of the exchanged particle or effective trajectory.  For example,
the $\gamma \gamma \to W^+ W^-$ differential cross section is
constant at high energies since the spin of the exchanged $W$ is
$\alpha_R = j = 1.$ In fact, after integration over phase space,
the cross section for pairs of vector bosons in photon-photon
collisions increases logarithmically with energy.  This is in
contrast to $\sigma(e^+ e^- \to W^+ W^-)$ which produces a single
$W^\pm$ pair in one partial wave and falls as $1/s.$ It is also
interesting that a dominant two-jet high $p_T$ reaction in
photon-photon collisions at high energies $s \gg p^2_T$ is $\gamma
\gamma \to g g$ which proceeds via two quark loops coupling via
gluon exchange in the $t$ channel\cite{HwangSjb}.

\section{Standard Model Tests}

Since each photon can be resolved into a $W^+ W^-$ pair, high
energy photon-photon collisions produce equivalent effective
$W^\pm$ beams, thus providing a remarkably background-free
laboratory for studying $W W$ interactions and testing for any
anomalous magnetic and quadrupole couplings.  The interacting
vector bosons can scatter pair-wise or annihilate; {\em e.g.},
they can annihilate into a Standard Model Higgs boson or a pair of
top quarks. There is thus a large array of tests of electroweak
theory possible in photon-photon collisions.  The splitting
function for $\gamma\to W^+ W^-$ can be relatively flat for some
$W$ helicities, so that one has a high probability for the $W$'s
to scatter with a high fraction of the energy of the photon. One
can thus study tree graphs contributions derived from photon, Z,
or Higgs exchange in the $t$-channel, and in the case of identical
$W$'s, the additional u-channel amplitudes.  In the case of
oppositely-charged $W$'s, $s$-channel annihilation processes such
as $W^+ W^- \to t\bar t$ contribute.  The largest cross sections
will arise if the $W$'s obey a strongly coupled theory; in this
case the longitudinal $W$'s scattering amplitude saturates
unitarity and the corresponding $\gamma\gamma \rightarrow WWWW$
cross section will be maximal. The cross sections of many Standard
Model processes are illustrated in Fig. 2. Reviews of this physics
are given in the
references.~\cite{Boos:2000ki,Asner:2001vh,Brodsky:1993xp,Chanowitz:1994aq,Gunion:1992ce}

\begin{figure}[htbp]
\begin{center}
\includegraphics[height=7in,width=6in]{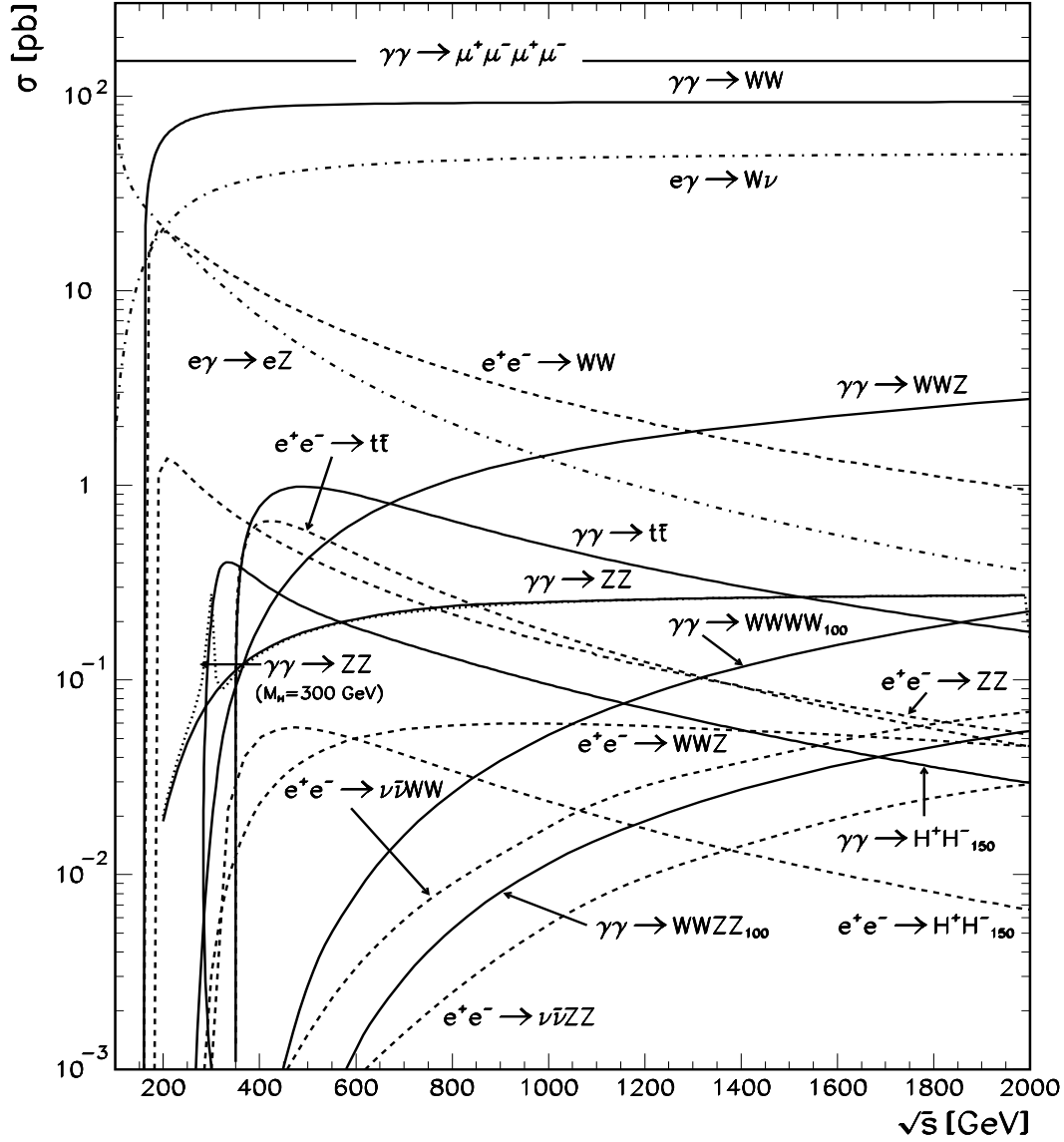}
\end{center}
\caption[*]{ Typical (unpolarized) cross sections in
$\gamma\gamma$, $\gamma e$ and $e^+e^-$ collisions. Solid,
dash-dotted and dashed curves correspond to $\gamma\gamma$,
$\gamma e$ and $e^+e^-$ modes respectively.  Unless indicated
otherwise the neutral Higgs mass was taken to be 100~GeV.  For
charged Higgs pair production, $M_{H^\pm}=150$~GeV was assumed.
From Boos {\em et al.}~\cite{Boos:2000ki}. \label{fig:cs}}
\end{figure}

One of the most important applications of two photon physics is
the direct production of $W$ pairs.  By using polarized
back-scattered laser beams, one can in principle study $\gamma
\gamma \to W^+ W^-$ production as a function of initial photon
helicities as well as resolve the $W$ helicities through their
decays.  The study of $\gamma \gamma \to W^+W^-$ is complimentary
to the corresponding $e^+ e^- \rightarrow W^+W^-$ channel, but it
also can check for the presence of anomalous four-point $\gamma
\gamma \to WW$ interactions not already constrained by
electromagnetic gauge invariance, such as the effects due to
$W^\ast$ exchange.

A main focus of the pair production measurements are the values of the
$W$ magnetic moment $ \mu_W = {e\over 2m_W}\ (1-\kappa-\lambda)$
and quadrupole moment $Q_W= - {e\over M^2_W}\ (\kappa-\lambda).$
The Standard Model predicts $\kappa=1$ and $\lambda=0,$ up
to radiative corrections analogous to the Schwinger corrections
to the electron anomalous moment.  The
anomalous moments are thus defined as $\mu_A = \mu_W-{e\over M_W}$
and $Q_A = Q_W + {e\over M^2_W}.$

The fact that $\mu_A$ and $Q_A$ are close to zero is actually a
general property of any  spin-one system if its size is small
compared to its Compton scale.  For example, consider the
Drell-Hearn-Gerasimov sum
rule~\cite{Drell:1966jv,Gerasimov:1965et} for the $W$ magnetic
moment: $ \mu^2_A = \left(\mu-{e\over M}\right)^2 =
{1\over\pi}\int^\infty _{\nu_{th}} {d\nu\over\nu}\,
[\sigma_P(\nu)-\sigma_A(\nu)]. $ Here $\sigma_{P(A)}$ is the total
photoabsorption cross section for photons on a $W$ with (anti-)
parallel helicities. As the radius of the $W$ becomes small, or
its threshold energy for inelastic excitation becomes large, the
DHG integral and hence $\mu^2_A$ vanishes. Hiller and I have shown
~\cite{Brodsky:1992px} that this argument can be generalized to
the spin-one anomalous quadrupole moment as well, by considering
one of the unsubtracted dispersion relations for near-forward
$\gamma$ spin-one Compton scattering~\cite{Tung:kn}:
\begin{eqnarray}
&& \mu_A^2 + {2t\over M^2_W}\  \left(\mu_A+{M_2\over W}\
Q_A\right)^2 = \nonumber \\
&& {1\over 4\pi} \int^\infty_{\nu_{th}}
{d\nu^2\over (\nu-t/4)^3}\ Im\, (f_P(s,t)-f_A(s,t))\ .
\end{eqnarray}
Here $\nu = (s-u)/4$.  One again sees that in the point-like
or high threshold energy limit, both
 $\mu_A \rightarrow 0,$ and
$Q_A\rightarrow 0.$ This result applies to any spin-one system,
even to the deuteron or the $\rho.$ The essential assumption is
the existence of the unsubtracted dispersion relations; {\em i.e.}, that
the anomalous moments are in principle computable quantities.

In the case of the $W$, the finite size correction is expected to
be order $m^2/\Lambda^2$, since the underlying composite theory
should be chiral to keep the $W$ mass finite as the composite
scale $\Lambda$ becomes large~\cite{Brodsky:1980zm}. Thus the fact
that a spin-one system has nearly  canonical values for its
moments signals that it has a small internal size; however, it
does not necessarily imply that it is a gauge field.

Yehudai~\cite{Yehudai:1991az} has made extensive studies of the
effect of anomalous moments on different helicity amplitude
contributing to $\gamma\gamma \rightarrow W^+W^-$ cross section.
The empirical sensitivity to anomalous couplings from
$\gamma\gamma$ reactions is comparable and complimentary to that
of $e^+ e^- \rightarrow W^+W^-.$  A comprehensive analysis of
fermion processes in photon-photon collisions is given by Layssac
and Renard~\cite{Layssac:2001ur}.

As emphasized by Jikia~ and Tkabladze~\cite{Jikia:1993pg}, pairs
of neutral gauge bosons can be produced in $\gamma\gamma$
reactions through one loop amplitudes in the Standard Model at a
rate which should be accessible to the NLC. Leptons, quarks, and
$W$ all contribute to the box graphs. The fermion and spin-one
exchange contributions to the $\gamma \gamma \to \gamma \gamma$
scattering amplitude have the characteristic behavior ${\cal M}
\sim s^0 f(t)$ and ${\cal M} \sim i\,s f(t)$ respectively.  The
latter is the dominant contribution at high energies, so one can
use the optical theorem to relate the forward imaginary part of
the scattering amplitude to the total $\gamma \gamma \to W^+ W^-$
cross section.  The resulting cross section $\sigma(\gamma\gamma
\rightarrow \gamma\gamma)$ is of order 20 fb at $\sqrt s_{\gamma
\gamma}$ , corresponding to 200 events/year at an NLC with
luminosity 10 fb$^{-1}$~\cite{Boos:2000ki}. The corresponding
$\gamma \gamma \to H^0 Z^0$ process has been analyzed by Gounaris,
Porfyriadis and Renard~\cite{Gounaris:2001rk}.

\begin{figure}[htb]
\begin{center}
\includegraphics[width=10cm,height=9cm]{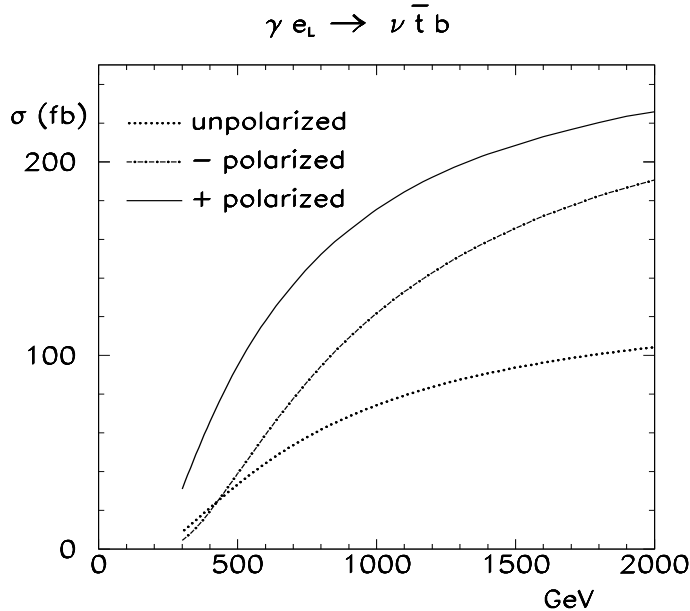}
\end{center}
\caption[*]{Single top quark production cross section in $\gamma
e$ collisions. From Boos {\em et al.}~\cite{Boos:2001sj}
\label{fig:f9}}
\end{figure}

A single top quark can be produced in electron-photon collisions
at an NLC through the process $e^-\gamma \rightarrow W^-
t\nu$~\cite{Jikia:1991hc}. See
Fig.~\ref{fig:f9}.~\cite{Boos:2001sj} This process can be
identified through the $t \to W^+ b$ decay with $W\ \to \ell \bar
\nu.$ The rate is strongly polarization dependent and is sensitive
to the structure of the $V_{tb}$ matrix element, possible fourth
generation quarks, and anomalous couplings. An interesting
background is the virtual $W$ process $e\gamma\rightarrow W^\ast
-\nu \to W^- H \nu,$ where the Higgs boson decays to $b\bar b$ and
$W^-\to \ell \bar \nu.$

\begin{figure}[htb]
\begin{center}
\includegraphics[height=3.truein]{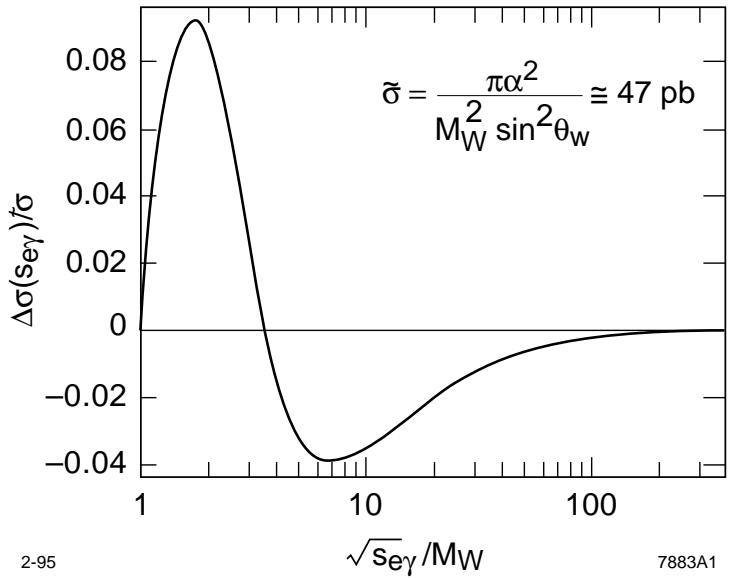}
\end{center}
\caption[*]{The Born cross section difference $\Delta \sigma$ for
the Standard Model process $\gamma e \to W \nu$ for parallel minus
antiparallel electron/photon helicities as a function of $\log
\sqrt s_{e \gamma}/M_W$ The logarithmic integral of $\Delta
\sigma$ vanishes in the classical limit.~\cite{Brodsky:1995ga}
\label{figB}}
\end{figure}

Schmidt, Rizzo, and I~\cite{Brodsky:1995ga,Rizzo:1999xj} have
shown that one can use the sign change of the integrand of the DHG
sum rule to test the canonical couplings of the Standard Model and
to isolate the higher order radiative corrections. For example,
consider the reactions $\gamma \gamma \to q \overline q$, $\gamma
e \to W \nu$ and $\gamma e \to Z e$ which can be studied with
back-scattered laser beams.  In contrast to the time-like process
$e^+ e^- \to W^+ W^-$, the $\gamma \gamma$ and $\gamma e$
reactions are sensitive to the anomalous moments of the gauge
bosons at $q^2 = 0.$ The vanishing of the logarithmic integral of
$\Delta \sigma$ in the Born approximation implies that there must
be a center-of-mass energy, $\sqrt s_0$, where the polarization
asymmetry $A=\Delta \sigma/ \sigma$ possesses a zero, {\em i.e.},
where $\Delta \sigma({\gamma e \to W \nu })$ reverses sign. The
cancellation of the positive and negative contributions
\cite{ginz} of $\Delta \sigma(\gamma e \to W \nu)$ to the DHG
integral is evident in Fig.~\ref{figB}. We find strong sensitivity
of the position of this zero or ``crossing point'' (which occurs
at $\sqrt s_{\gamma e} = 3.1583 \ldots  M_W \simeq 254$ GeV in the
SM) to modifications of the SM trilinear $\gamma W W$ coupling and
thus can lead to high precision constraints.  In addition to the
fact that only a limited range of energy is required, the
polarization asymmetry measurements have the advantage that many
of the systematic errors cancel in taking cross section ratios.
This technique can clearly be generalized to other higher order
tree-graph processes in the Standard Model and supersymmetric
gauge theory.  The position of the zero in the photoabsorption
asymmetry thus provides an additional weapon in the arsenal used
to probe anomalous trilinear gauge couplings.

\vspace{.5cm}

\section{Inclusive QCD Tests}

Because of the simplicity of its initial state, two-photon
collisions provide an important laboratory for testing coherent
and incoherent effects in quantum chromodynamics. In QCD events
where each photon is resolved~\cite{Brodsky:1978rp,Drees:1995ti}
in terms of its intermediate quark and gluon states,  $\gamma
\gamma$ collisions resemble point-like meson-meson collisions. One
can study detailed features of $\gamma \gamma \to t \bar t$ at
threshold and its final state evolution.  In the case of single or
double diffractive two-photon events, one can study fundamental
aspects of pomeron and odderon $t$-channel physics.  For example,
the energy asymmetry of charm quark versus anti-charm jets in
four-jet reactions can measure the interference of the pomeron and
odderon~\cite{Brodsky:1999mz}.

Consider, the cross section for producing two quark pairs. $\gamma
\gamma \to q_1 \bar q_1 q_2 \bar q_2$, which can be described as
the interaction of two color
dipoles~\cite{Brodsky:1997sd,Bartels:1996ke}.  If one of the heavy
quark pairs is a charm or bottom quark pair with a small color
dipole moment, then the multiple gluonic exchange graphs will
Reggeize into a hard pomeron, and one predict a strong energy
growth of the production cross section similar to that observed at
HERA in diffractive charm electroproduction~\cite{Lipatov:2000me}.
The QCD description of the hard QCD pomeron is given by the BFKL
analysis~\cite{Fadin:1998py}.  There has recently been progress in
stabilizing the BFKL predictions at next-to-leading-order by using
BLM scale fixing.~\cite{Brodsky:1998kn}

\section{The Photon Structure Functions}

One can also utilize electron-photon collisions at a linear
collider to test the shape and growth of the photon structure
functions~\cite{Brodsky:1971vm,Walsh:xy,Krawczyk:2000nh,Drees:1995ti}
The back-scattered laser beam provides a high energy polarized
target photon, and the neutral current probe is obtained by
tagging the scattered electron at momentum transfer squared $Q^2.$
One can also reconstruct the charged current contributions where
the electron scatters into a neutrino from calorimetric
measurements of the recoiling system. It also should be possible
to identify the separate charm, bottom, top and $W$ contributions
to the photon structure functions.

The photon structure functions receive hadron-like contributions
from the photon's resolved Fock components as well as its direct
component derived from the $\gamma^\ast \gamma \to q \bar q$~
time-like QCD Compton amplitude. Because of the direct
contributions, the photon structure functions obey an
inhomogeneous evolution equation. The result,  as first shown by
Witten~\cite{Witten:ju} is that the leading order QCD structure
functions of the photon have a unique scaling behavior:
$F_1(x,Q^2) = h(x)\, \ell n {Q^2\over \Lambda^2}$, $F_2(x,Q^2) =
f_2(x)$, $F_3(x,Q^2) = f^{\rm Box}_3(x)$.

The most characteristic behavior of the photon structure function
$F_2^\gamma(x,Q^2)$ in QCD is its continuous linear rise of with
$\log Q^2$ at fixed $x$.  As emphasized by Peterson, Walsh and
Zerwas~\cite{Peterson:1982tt}, the fact that this tree graph
behavior is preserved to all orders in perturbation theory is due
to the balance in QCD between the increase of the phase space for
gluon emission in the scattering processes versus the decreasing
strength of the gluon coupling due to asymptotic freedom. Although
the logarithmic rise of the Born approximation result is
preserved, the shape of $h(x)$ is modified by the QCD radiation.
If the running coupling constant were to freeze to a constant
value at large momentum transfer, the photon structure function
stops rising at high $Q^2$ due to the increased phase space for
gluon radiation. Thus probing the QCD photon structure functions
at the high momentum transfers available at the NLC will provide a
valuable test of asymptotic freedom.

\section{The Photon Structure Function and Final-State Interactions}

Recently, Hoyer, Marchal, Peigne, and Sannino and
I~\cite{Brodsky:2001ue} have challenged the common view that
structure functions measured in deep inelastic lepton scattering
are determined by the probability of finding quarks and gluons in
the target hadron.  We show that this is not correct in gauge
theory. Gluon exchange between the fast, outgoing partons and
target spectators, which is usually assumed to be an irrelevant
gauge artifact, affects the leading twist structure functions in a
profound way.  This observation removes the apparent contradiction
between the projectile (eikonal) and target (parton model) views
of diffractive and small $x_{\rm Bjorken}$ phenomena.  The
diffractive scattering of the fast outgoing quarks on spectators
in the target in turn causes shadowing in the DIS cross section.
Thus the depletion of the nuclear structure functions is not
intrinsic to the wave function of the nucleus, but is a coherent
effect arising from the destructive interference of diffractive
channels induced by final-state interactions.  This is consistent
with the Glauber-Gribov interpretation of shadowing as a
rescattering effect.  Similar effects can be present in the photon
structure function; {\em i.e.}, the photon structure function is
modified by rescattering of the struck quark with the photon's
spectator system.

\section{Single-Spin Asymmetries in Photon-Photon Collisions}

Measurements from the HERMES and SMC collaborations show a
remarkably large azimuthal single-spin asymmetries $A_{UL}$ and
$A_{UT}$ of the proton in semi-inclusive pion leptoproduction
$\gamma^*(q) p \to \pi X.$ Recently, Dae Sung Hwang and Ivan
Schmidt and I~\cite{Brodsky:2002cx} have shown that final-state
interactions from gluon exchange between the outgoing quark and
the target spectator system lead to single-spin asymmetries in
deep inelastic lepton-proton scattering at leading twist in
perturbative QCD; {\em i.e.}, the rescattering corrections are not
power-law suppressed at large photon virtuality $Q^2$ at fixed
$x_{bj}$.  The existence of such single-spin asymmetries requires
a phase difference between two amplitudes coupling the proton
target with $J^z_p = \pm {1\over 2}$ to the same final-state, the
same amplitudes which are necessary to produce a nonzero proton
anomalous magnetic moment.  We have shown that the exchange of
gauge particles between the outgoing quark and the proton
spectators produces a Coulomb-like complex phase which depends on
the angular momentum $L^z$ of the proton's constituents and is
thus distinct for different proton spin amplitudes.  The
single-spin asymmetry which arises from such final-state
interactions does not factorize into a product of distribution
function and fragmentation function, and it is not related to the
transversity distribution $\delta q(x,Q)$ which correlates
transversely polarized quarks with the spin of the transversely
polarized target nucleon.  These effects highlight the unexpected
importance of final and initial state interactions in QCD
observables.

Final state interactions will also lead to new types of single
spin asymmetries in photon-photon collisions. For example, in
$\gamma^* \gamma \to \pi X$ and $\gamma^* \gamma \to {\rm jet} X$
we expect $T$-odd correlations of the type $\vec S_\gamma \cdot
\vec q \times \vec p$ where $\vec S_\gamma$ is the polarization of
the real photon, $\vec q$ is the beam direction of an incident
virtual photon, and $\vec p$ is the direction of a produced quark
or hadron.  The resulting asymmetry of the photon polarized normal
to the production plane will be leading twist.  As in the proton
target case, the single-spin asymmetry will be sensitive to
orbital angular momentum in the photon wavefunction and details of
the photon structure at the amplitude level.

\section{Single and Double Diffraction in Photon-Photon Collisions}

The high energies of a photon-photon collider will make the study
of double diffractive $\gamma \gamma \to V^0 V^0$ and
semi-inclusive single diffractive processes $\gamma \gamma \to V^0
X$ in the Regge regime $s >> |t|$ interesting. Here $V^0 = \rho,
\omega\phi,J/\psi,\cdots$ If $|t|$ is taken larger than the QCD
confinement scale, then one has the potential for a detailed study
of fundamental Pomeron processes and its gluonic composition. As
in the case of large angle exclusive $\gamma \gamma$ processes,
the scattering amplitude is computed by convoluting the hard
scattering PQCD amplitude for $\gamma \gamma \to q \bar q q \bar
q$ with the vector meson distribution amplitudes.  The two gluon
exchange contribution dominates in the Regge
regime~\cite{Chernyak:xe}, giving a characteristic exclusive
process scaling law of order $ {d\sigma\over dt}\
(\gamma\gamma\rightarrow V^0V^0)\sim {\alpha^4_s(t) / t^6}.$
Ginzburg, Ivanov and Serbo~\cite{Ginzburg:gy} have emphasized that
the corresponding $\gamma\gamma\rightarrow $ pseudoscalar and
tensor meson channels can be used to isolate the Odderon exchange
contribution, contributions related at a fundamental level to
three gluon exchange.

In addition, the photon can diffractively dissociate into quark
pairs $\gamma e \to q \bar q e'$ by Coulomb scattering on the
incoming electron.  This measures the transverse derivative of the
photon wavefunction ${\partial \over \partial k_\perp} \psi_{q
\bar q}(x,k_\perp,\lambda_i).$ This is the analog of the E791
experiment at Fermilab~\cite{Ashery:1999nq} which resolved the
pion light-front wavefunction by diffractive dissociation $\pi A
\to q \bar q A'$ on a nuclear target.  The results of the
diffractive pion experiment are consistent with color
transparency, and the momentum partition of the jets conforms
closely with the shape of the asymptotic distribution amplitude,
$\phi^\mathrm{asympt}_\pi (x) = \sqrt 3 f_\pi x(1-x)$,
corresponding to the leading anomalous dimension solution
\cite{BrodskyLepage} to the perturbative QCD evolution equation.

\section{Other QCD Tests in Photon-Photon Collisions}

Two-photon annihilation $\gamma^*(q_1) \gamma^*(q_2) \to $ hadrons
for real and virtual photons can thus provide some of the most detailed
and incisive tests of QCD.  Among the processes of special interest
are:

\begin{enumerate}

\item the production of four jets such as $\gamma \gamma \to c \bar c c
\bar c$ can test Fermi-color statistics for charm quarks by checking for
the interference effects of like sign quarks~\cite{BA}.

\item the total two-photon annihilation hadronic cross section
$\sigma(s,q^2_1,q^2_2),$ which is related to the light-by-light
hadronic contribution to the muon anomalous moment;

\item the formation of $C = +$ hadronic resonances, which can
reveal
 exotic states such as $q
\bar q g$ hybrids and discriminate gluonium formation
\cite{Pennington:2000ai,Acciarri:2001ex}.  The production of the
$\eta_B$ and $\chi_B$ states are essentially unexplored in
QCD.~\cite{Heister:2002if}

\item single-hadron processes such as $\gamma^* \gamma^* \to
\pi^0$, which test the transition from the anomaly-dominated pion
decay constant to the short-distance structure of currents
dictated by the operator-product expansion and perturbative QCD
factorization theorems;

\item hadron pair production processes such as $\gamma^* \gamma \to
\pi^+ \pi^-, K^+ K^-, p \bar p$, which at fixed invariant pair
mass measures the $s \to t$ crossing of the virtual Compton
amplitude \cite{Brodsky:1981rp,BrodskyLepage}. I discuss this
further in the next section. When one photon is highly virtual,
these exclusive hadron production channels are dual to the photon
structure function $F^\gamma_2(x,Q^2)$ in the endpoint $x \to 1$
region at fixed invariant pair mass.  The leading twist-amplitude
for $\gamma^* \gamma \to \pi^+ \pi^-$ is sensitive to the $1/x -
1/(1-x)$ moment of the $q \bar q$ distribution amplitude
$\Phi_{\pi^+ \pi^-}(x,Q^2)$ of the two-pion system
\cite{Muller:1994fv,Diehl:2000uv}, the time-like extension of
skewed parton distributions.  In addition one can measure the pion
charge asymmetry in $e^+ e^- \to \pi^+ \pi^- e^+ e^-$ arising from
the interference of the $\gamma \gamma \to \pi^+ \pi^-$ Compton
amplitude with the time-like pion form factor
\cite{Brodsky:1971ud}.  At the unphysical point $s =q^2_1= q^2_2 =
0$, the amplitude is fixed by the low energy theorem to the hadron
charge squared.  The ratio of the measured $\gamma \gamma \to
\Lambda \bar \Lambda$ and $\gamma \gamma \to p \bar p$ cross
sections is anomalous at threshold, a fact which may be associated
with the soliton structure of baryons in
QCD~\cite{Sommermann:1992yh,Karliner:2002nk};

\item Exclusive or semi-inclusive channels can also be
studied by coalescence of the produced quarks.  An interesting
example is higher generation final state such as $\gamma \gamma
\to B_c \bar B_c, $ which can have very complex angular
structure~\cite{Brodsky:1985cr}.

\item As pointed out by Hwang~\cite{Hwang}, one can study deeply virtual Compton
scattering on a photon target in $e \gamma$ collisions to
determine the light-cone wavefunctions and other features of the
photon~\cite{Brodsky:2000xy}.

\end{enumerate}

\section{The Photon-to-Pion Transition Form Factor and the Pion
Distribution Amplitude}

The simplest and perhaps most elegant illustration of an exclusive
reaction in QCD is the evaluation of the photon-to-pion transition
form factor $F_{\gamma \to \pi}(Q^2)$ which is measurable in
single-tagged two-photon $ee \to ee \pi^0$ reactions.  The form
factor is defined via the invariant amplitude $\Gamma^\mu = -ie^2
F_{\pi \gamma}(Q^2) \varepsilon^{\mu \nu \rho \sigma} p^\pi_\nu
\varepsilon_\rho q_\sigma$.  As in inclusive reactions, one must
specify a factorization scheme which divides the integration
regions of the loop integrals into hard and soft momenta, compared
to the resolution scale $\widetilde Q$.  At leading twist, the
transition form factor then factorizes as a convolution of the
$\gamma^* \gamma \to q \bar q$ amplitude (where the quarks are
collinear with the final state pion) with the valence light-cone
wavefunction of the pion \cite{BrodskyLepage}:
\begin{equation}
F_{\gamma M}(Q^2)= {4 \over \sqrt 3}\int^1_0 dx
\phi_M(x,\widetilde Q) T^H_{\gamma \to M}(x,Q^2) .
\label{transitionformfactor}
\end{equation}
The hard scattering amplitude for $\gamma\gamma^*\to q \bar q$ is
$ T^H_{\gamma M}(x,Q^2) = { [(1-x) Q^2]^{-1}}\left(1 + {\cal
O}(\alpha_s)\right)$.  For the asymptotic distribution amplitude
$\phi^\mathrm{asympt}_\pi (x) = \sqrt 3 f_\pi x(1-x)$ one predicts
\cite{Brodsky:1997dh}
\begin{displaymath}
Q^2 F_{\gamma \pi}(Q^2)= 2 f_\pi \left(1 - {5\over 3}
{\alpha_V(Q^*)\over \pi}\right)
\end{displaymath}
where $Q^*= e^{-3/2} Q$ is the estimated BLM scale for the pion
form factor in the $V$ scheme.

\begin{figure}[htbp]
\centering
\includegraphics[width=.69\columnwidth]{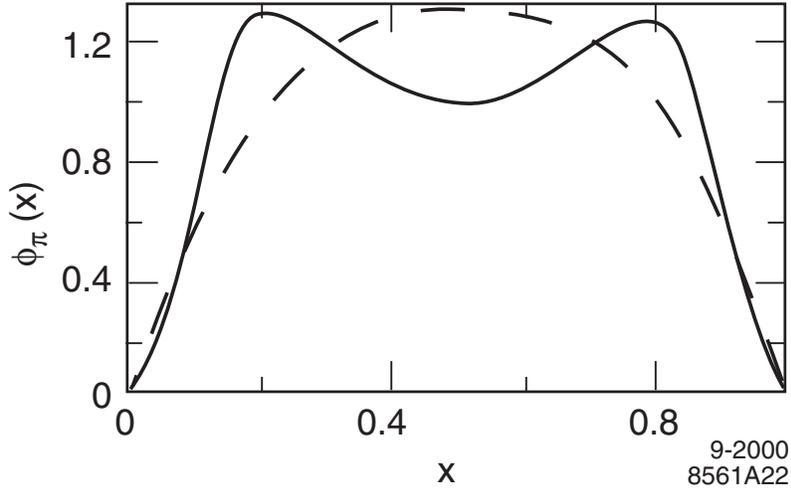}\\
(a)\\[15pt]
\includegraphics[width=.69\columnwidth]{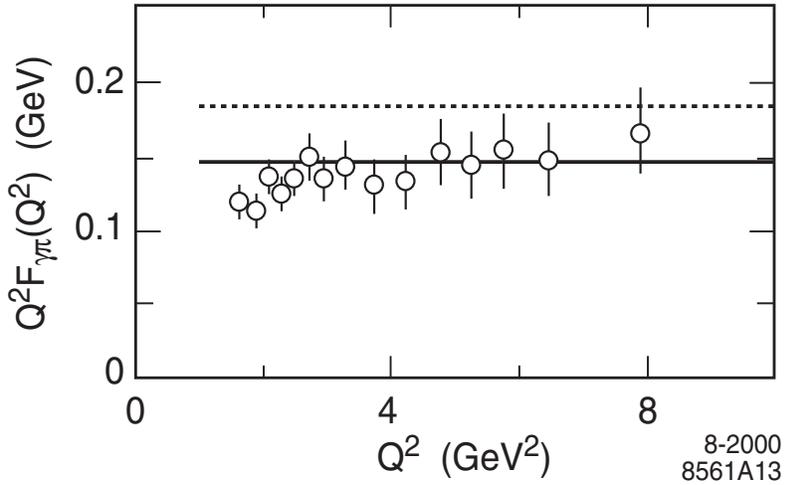}\\
(b) \caption[*]{ (a) Transverse lattice results for the pion
distribution amplitude at $Q^2 \sim 10 {\rm GeV}^2$.  The solid
curve is the theoretical prediction from the combined
DLCQ/transverse lattice method \cite{Dalley:2000dh}; the chain
line is the experimental result obtained from di-jet diffractive
dissociation \cite{Ashery:1999nq,Aitala:2001hb}.  Both are
normalized to the same area for comparison.  (b) Scaling of the
transition photon to pion transition form factor $Q^2F_{\gamma
\pi^0}(Q^2)$.  The dotted and solid theoretical curves are the
perturbative QCD prediction at leading and next-to-leading order,
respectively, assuming the asymptotic pion distribution The data
are from the CLEO collaboration \cite{Gronberg:1998fj}.
\label{Fig:DalleyCleo}}
\end{figure}

The PQCD predictions have been tested in measurements of $e \gamma
\to e \pi^0$ by the CLEO collaboration \cite{Gronberg:1998fj} (see
Fig. \ref{Fig:DalleyCleo} (b)).  The flat scaling of the $Q^2
F_{\gamma \pi}(Q^2)$ data from $Q^2 = 2$ to $Q^2 = 8$ GeV$^2$
provides an important confirmation of the applicability of leading
twist QCD to this process.  The magnitude of $Q^2 F_{\gamma
\pi}(Q^2)$ is remarkably consistent with the predicted form,
assuming the asymptotic distribution amplitude and including the
LO QCD radiative correction with $\alpha_V(e^{-3/2} Q)/\pi \simeq
0.12$.  One could allow for some broadening of the distribution
amplitude with a corresponding increase in the value of $\alpha_V$
at small scales.  Radyushkin \cite{Radyushkin}, Ong \cite{Ong} and
Kroll \cite{Kroll} have also noted that the scaling and
normalization of the photon to pion transition form factor tends
to favor the asymptotic form for the pion distribution amplitude
and rules out broader distributions such as the two-humped form
suggested by QCD sum rules \cite{CZ}.  More detailed studies are
given in the
references.~\cite{Khodjamirian:1999tk,Schmedding:2000ap,Bakulev:2001pa}

When both photons are virtual, the denominator of $T_H$ for the
$\gamma \gamma^* \to \pi^0$ reaction becomes $(1-x) Q^2_1 + x
Q^2_2$ \cite{BrodskyLepage,Ong}, and the amplitude becomes nearly
insensitive to the shape of the distribution amplitude once it is
normalized to the pion decay constant.  Thus the ratio of singly
virtual to doubly virtual pion production is particularly
sensitive to the shape of $\phi_\pi(x,Q^2)$ since higher order
corrections and normalization errors tend to cancel in the ratio.

\null

\section{Exclusive Two-Photon Annihilation into Hadron Pairs}

At large momentum transfer, the angular distribution of hadron
pairs produced by photon-photon annihilation are among the best
determinants of the shape of the meson and baryon distribution
amplitudes $\phi_M(x,Q)$ and $\phi_B(x_i,Q),$ which control almost
all exclusive processes involving a hard scale $Q$. The
determination of the shape and normalization of the distribution
amplitudes, which are gauge-invariant and process-independent
measures of the valence wavefunctions of the hadrons, has become
particularly important in view of their importance in the analysis
of exclusive semi-leptonic and two-body hadronic $B$-decays
\cite{BHS,Sz,Beneke:1999br,Keum:2000ph,Keum:2000wi}.  There has
also been considerable progress both in calculating hadron
wavefunctions from first principles in QCD and in measuring them
using diffractive di-jet dissociation.

Much of this important two-photon physics is also accessible at
low energy, high luminosity $e^+ e^-$ colliders, particularly for
measurements of channels important in the light-by-light
contribution to the muon $g$--2 and the exploration of the
transition between threshold amplitudes which are controlled by
low-energy effective theories such as the chiral Hamiltonian
through the transition to the domain where leading-twist
perturbative QCD becomes applicable.  There have been almost no
measurements of double-tagged events needed to unravel the
separate $q^2_1$ and $q^2_2$ dependence of photon-photon
annihilation.  Hadron pair production from two-photon annihilation
plays a crucial role in unravelling the perturbative and
non-perturbative structure of QCD, first by testing the validity
and empirical applicability of leading-twist factorization
theorems, second by verifying the structure of the underlying
perturbative QCD subprocesses, and third, through measurements of
angular distributions and ratios which are sensitive to the shape
of the distribution amplitudes.  In effect, photon-photon
collisions provide a microscope for testing fundamental scaling
laws of PQCD and for measuring distribution amplitudes.

Two-photon reactions, $\gamma \gamma \to H \bar H$ at large $s =
(k_1 + k_2)^2$ and fixed $\theta_\mathrm{cm}$, provide a
particularly important laboratory for testing QCD since these
cross-channel ``Compton'' processes are the simplest calculable
large-angle exclusive hadronic scattering reactions.  The helicity
structure, and often even the absolute normalization can be
rigorously computed for each two-photon channel
\cite{Brodsky:1981rp}.  In the case of meson pairs, dimensional
counting predicts that for large $s$, $s^4 d\sigma/dt(\gamma
\gamma \to M \bar M)$ scales at fixed $t/s$ or
$\theta_\mathrm{cm}$ up to factors of $\ln s/\Lambda^2$.  The
angular dependence of the $\gamma \gamma \to H \bar H$ amplitudes
can be used to determine the shape of the process-independent
distribution amplitudes, $\phi_H(x,Q)$.  An important feature of
the $\gamma \gamma \to M \bar M$ amplitude for meson pairs is that
the contributions of Landshoff pitch singularities are power-law
suppressed at the Born level---even before taking into account
Sudakov form factor suppression.  There are also no anomalous
contributions from the $x \to 1$ endpoint integration region.
Thus, as in the calculation of the meson form factors, each
fixed-angle helicity amplitude can be written to leading order in
$1/Q$ in the factorized form $[Q^2 = p_T^2 = tu/s$; $\widetilde
Q_x = \min(xQ,(l-x)Q)]$:
\begin{eqnarray}
\mathcal{M}_{\gamma \gamma\to M \bar M} &= &\int^1_0\, dx \int^1_0
\, dy \nonumber\\
&&\hspace{-2pc}\phi_{\bar M}(y,\widetilde Q_y)
T_H(x,y,s,\theta_\mathrm{cm} \phi_{M}(x,\widetilde Q_x) ,
\end{eqnarray}
where $T_H$ is the hard-scattering amplitude $\gamma \gamma \to (q
\bar q) (q \bar q)$ for the production of the valence quarks
collinear with each meson, and $\phi_M(x,\widetilde Q)$ is the
amplitude for finding the valence $q$ and $\bar q$ with light-cone
fractions of the meson's momentum, integrated over transverse
momenta $k_\perp < \widetilde Q$.  The contribution of non-valence
Fock states are power-law suppressed.  Furthermore, the
helicity-selection rules \cite{Brodsky:1981kj} of perturbative QCD
predict that vector mesons are produced with opposite helicities
to leading order in $1/Q$ and all orders in $\alpha_s$.  The
dependence in $x$ and $y$ of several terms in $T_{\lambda,
\lambda'}$ is quite similar to that appearing in the meson's
electromagnetic form factor.  Thus much of the dependence on
$\phi_M(x,Q)$ can be eliminated by expressing it in terms of the
meson form factor.  In fact, the ratio of the $\gamma \gamma \to
\pi^+ \pi^-$ and $e^+ e^- \to \mu^+ \mu^-$ amplitudes at large $s$
and fixed $\theta_{CM}$ is nearly insensitive to the running
coupling and the shape of the pion distribution amplitude:
\begin{equation}{{d\sigma \over dt }(\gamma \gamma \to \pi^+ \pi^-)
\over {d\sigma \over dt }(\gamma \gamma \to \mu^+ \mu^-)} \sim {4
\vert F_\pi(s) \vert^2 \over 1 - \cos^2 \theta_\mathrm{cm} }.
\end{equation}
\begin{figure*}[htb]
\begin{center}
\includegraphics[width=.9\textwidth]{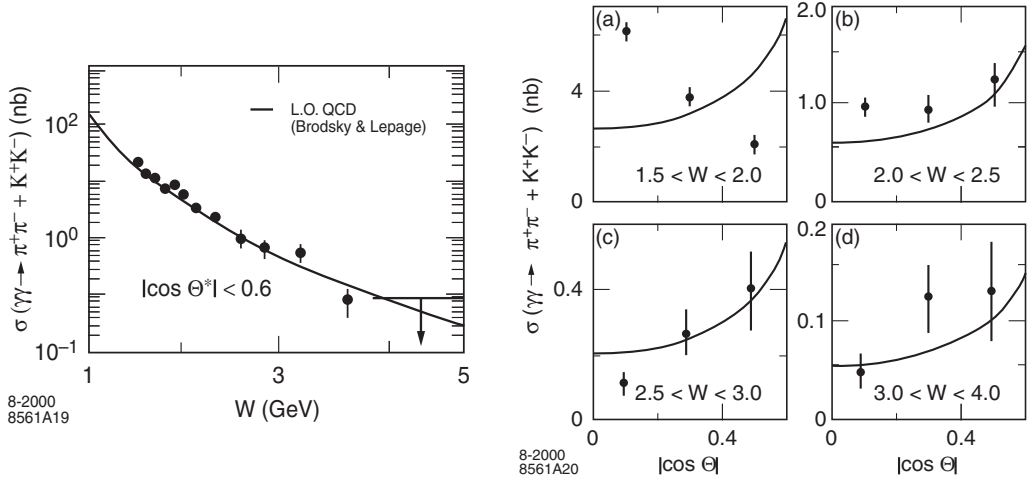}
\end{center}
\caption[*]{Comparison of the sum of $\gamma \gamma \rightarrow
\pi^+ \pi^-$ and $\gamma \gamma \rightarrow K^+ K^-$ meson pair
production cross sections with the scaling and angular
distribution of the perturbative QCD prediction
\cite{Brodsky:1981rp}.  The data are from the CLEO collaboration
\cite{Dominick:1994bw}. \label{Fig:CLEO}}
\end{figure*}
The comparison of the PQCD prediction for the sum of $\pi^+ \pi^-$
plus $K^+ K^-$ channels with CLEO data \cite{Dominick:1994bw} is
shown in Fig. \ref{Fig:CLEO}.  The CLEO data for charged pion and
kaon pairs show a clear transition to the scaling and angular
distribution predicted by PQCD \cite{Brodsky:1981rp} for $W =
\sqrt(s_{\gamma \gamma} > 2$ GeV. It is clearly important to
measure the magnitude and angular dependence of the two-photon
production of neutral pions and $\rho^+ \rho^-$ in view of the
strong sensitivity of these channels to the shape of meson
distribution amplitudes (see Figs. \ref{Fig:piangle} and
\ref{Fig:rhoangle}).  QCD also predicts that the production cross
section for charged $\rho$-pairs (with any helicity) is much
larger than for that of neutral $\rho$ pairs, particularly at
large $\theta_\mathrm{cm}$ angles.  Similar predictions are
possible for other helicity-zero mesons.  For an alternative model
based on the QCD ``handbag" diagram, see the
references.~\cite{Diehl:2001fv}

\begin{figure}[htbp]
\begin{center}
\includegraphics[width=.48\columnwidth]{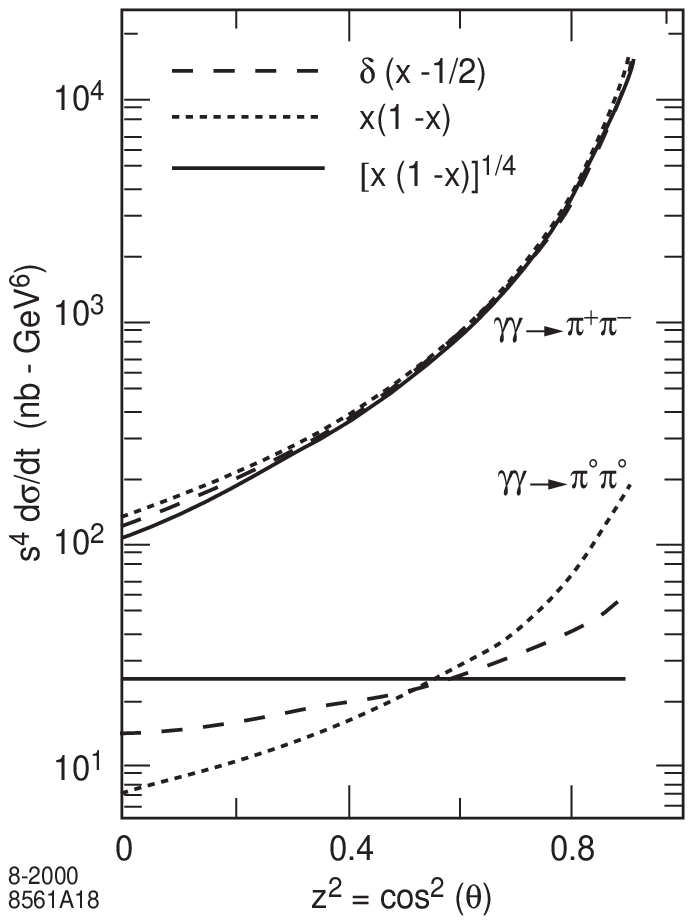}
\end{center}
\caption[*]{Predictions for the angular distribution of the
$\gamma\gamma\rightarrow \pi^+\pi^-$ and $\gamma \gamma
\rightarrow \pi^0 \pi^0$ pair production cross sections for three
different pion distribution amplitudes \cite{Brodsky:1981rp}.
\label{Fig:piangle}}
\bigskip
\begin{center}
\includegraphics[width=.48\columnwidth]{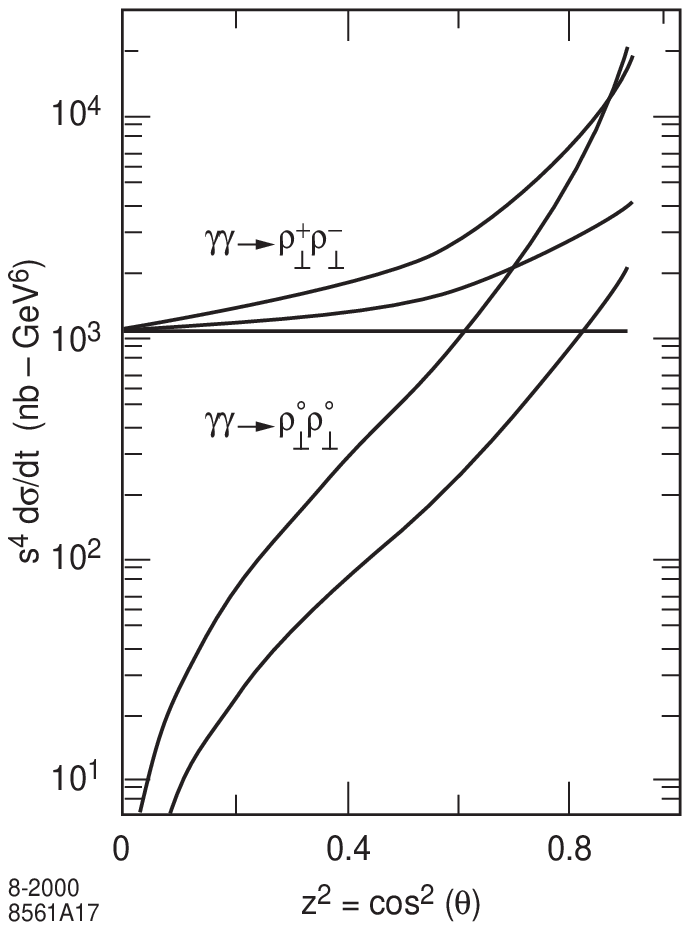}
\end{center}
\caption[*]{Predictions for the angular distribution of the
$\gamma\gamma\rightarrow \rho^+\rho^-$ and $\gamma \gamma
\rightarrow \rho^0 \rho^0$ pair production cross sections for
three different $\rho$ distribution amplitudes as in Fig.
\ref{Fig:piangle} \cite{Brodsky:1981rp}. \label{Fig:rhoangle}}
\end{figure}

\begin{figure}[!t]
\begin{center}
\includegraphics[width=5in]{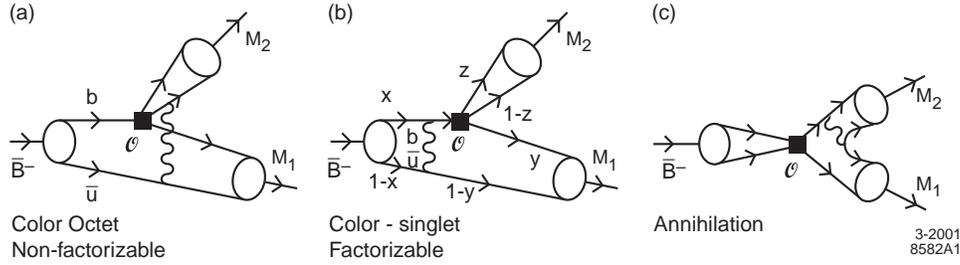}
\end{center}
\caption[*]{Three representative contributions to exclusive $B$
decays to meson pairs in PQCD.  The operators $\cal O$ represent
the QCD-improved effective weak interaction. \label{fig:B}}
\end{figure}

As noted above, the analysis of exclusive $B$ decays has much in
common with the analysis of exclusive two-photon reactions
\cite{Brodsky:2001jw}. For example, consider the three
representative contributions to the decay of a $B$ meson to meson
pairs illustrated in Fig. \ref{fig:B}.  In Fig. \ref{fig:B}(a) the
weak interaction effective operator $\mathcal{O}$ produces a $ q
\bar q$ in a color octet state.  A gluon with virtuality $Q^2 =
\mathcal{O} (M_B^2)$ is needed to equilibrate the large momentum
fraction carried by the $b$ quark in the $\bar B$ wavefunction.
The amplitude then factors into a hard QCD/electroweak subprocess
amplitude for quarks which are collinear with their respective
hadrons: $T_H([b(x) \bar u(1-x)] \to [q(y) \bar u(1-y)]_1 [q(z)
\bar q(1-z)]_2)$ convoluted with the distribution amplitudes
$\phi(x,Q)$ \cite{BrodskyLepage} of the incident and final
hadrons:
\begin{eqnarray*}
\mathcal{M}_\mathrm{octet}(B \to M_1 M_2) &= &\int^1_0\, dz
\int^1_0\, dy \int^1_0\,
dx\\
&&\hspace{-3pc}\phi_B(x,Q) T_H(x,y,z) \phi_{M_1}(y,Q)
\phi_{M_2}(z,Q).
\end{eqnarray*}
Here $x = k^{+}/p^{+}_H = (k^0+ k^z) /(p^0_H + p^z_H)$ are the
light-cone momentum fractions carried by the valence quarks.

There are a several features of QCD which are required to ensure
the consistency of the PQCD approach: (a) the effective QCD
coupling $\alpha_s(Q^2)$ needs to be under control at the relevant
scales of $B$ decay; (b) the distribution amplitudes of the
hadrons need to satisfy convergence properties at the endpoints;
and (c) one requires the coherent cancellation of the couplings of
soft gluons to color-singlet states.  This property, color
transparency \cite{Brodsky:1988xz}, is a fundamental coherence
property of gauge theory and leads to diminished final-state
interactions and corrections to the PQCD factorizable
contributions.  The problem of setting the renormalization scale
of the coupling for exclusive amplitudes is discussed in
\cite{Brodsky:1997dh}.  Conformal symmetry can be used as a guide
to organize the hard scattering calculation and to determine the
leading contributions to the hadron distribution
amplitudes~\cite{Brodsky:1980ny,Brodsky:2000cr,Braun:1999te}.

Baryon pair production in two-photon annihilation is also an
important testing ground for QCD.  The calculation of $T_H$ for
Compton scattering requires the evaluation of 368
helicity-conserving tree diagrams which contribute to $\gamma
(qqq) \to \gamma^\prime (qqq)^\prime$ at the Born level and a
careful integration over singular intermediate energy denominators
\cite{Farrar:1990qj,Kronfeld:1991kp,Guichon:1998xv}.  Brooks and
Dixon \cite{Brooks:2000nb} have completed a recalculation of the
proton Compton process at leading order in PQCD, extending and
correcting earlier work.  It is useful to consider the ratio $
d\sigma/dt(\gamma \gamma \to \bar p p)/ d\sigma/dt(e^+ e^- \to
\bar p p)$ since the power-law fall-off, the normalization of the
valence wavefunctions, and much of the uncertainty from the scale
of the QCD coupling cancel.  The scaling and angular dependence of
this ratio is sensitive to the shape of the proton distribution
amplitudes.  The perturbative QCD predictions for the phase of the
Compton amplitude phase can be tested in virtual Compton
scattering by interference with Bethe-Heitler processes
\cite{Brodsky:1972vv}.

It is also interesting to measure baryon and isobar pair
production in two photon reactions near threshold.  Ratios such as
$\sigma(\gamma \gamma \to \Delta^{++} \Delta^{--})/\sigma(\gamma
\gamma \to \Delta^{+} \Delta^{-})$ can be as large as $16:1$ in
the quark model since the three-quark wavefunction of the $\Delta$
is expected to be symmetric.  Such large ratios would not be
expected in soliton models \cite{Sommermann:1992yh} in which
intermediate multi-pion channels play a major role.

Pobylitsa {\em et al.} \cite{Pobylitsa:2001cz} have shown how the
predictions of perturbative QCD can be extended to processes such
as $\gamma \gamma \to p \bar p \pi$ where the pion is produced at
low velocities relative to that of the $p$ or $\bar p$ by
utilizing soft pion theorems in analogy to soft photon theorems in
QED.  The distribution amplitude of the $p \pi$ composite is
obtained from the proton distribution amplitude from a chiral
rotation.  A test of this procedure in inelastic electron
scattering at large momentum transfer $e p \to p \pi$ and small
invariant $p^\prime \pi$ mass has been remarkably successful.
Many tests of the soft meson procedure are possible in
multiparticle $e^+ e^-$ and $\gamma \gamma$ final states.

The leading-twist QCD predictions for exclusive two-photon
processes such as the photon-to-pion transition form factor and
$\gamma \gamma \to $ hadron pairs are based on rigorous
factorization theorems.  The data from the CLEO collaboration on
$F_{\gamma \pi}(Q^2)$ and the sum of $\gamma \gamma \to \pi^+
\pi^-$ and $\gamma \gamma \to K^+ K^-$ channels are in excellent
agreement with the QCD predictions.  It is particularly compelling
to see a transition in angular dependence between the low energy
chiral and PQCD regimes.  The success of leading-twist
perturbative QCD scaling for exclusive processes at presently
experimentally accessible momentum transfer can be understood if
the effective coupling $\alpha_V(Q^*)$ is approximately constant
at the relatively small scales $Q^*$ relevant to the hard
scattering amplitudes \cite{Brodsky:1997dh}. The evolution of the
quark distribution amplitudes in the low-$Q^*$ domain also needs
to be minimal.  Sudakov suppression of the endpoint contributions
is also strengthened if the coupling is frozen because of the
exponentiation of a double logarithmic series.

One of the formidable challenges in QCD is the calculation of
non-perturbative wavefunctions of hadrons from first principles.
The calculations of the pion distribution amplitude by Dalley
\cite{Dalley:2000dh} and by Burkardt and Seal
\cite{Burkardt:2001mf} using light-cone and transverse lattice
methods is particularly encouraging.  The predicted form of
$\phi_\pi(x,Q)$ is somewhat broader than but not inconsistent with
the asymptotic form favored by the measured normalization of $Q^2
F_{\gamma \pi^0}(Q^2)$ and the pion wavefunction inferred from
diffractive di-jet production.

Clearly much more experimental input on hadron wavefunctions is
needed, particularly from measurements of two-photon exclusive
reactions into meson and baryon pairs at the high luminosity $B$
factories.  For example, as shown in Fig. \ref{Fig:piangle}, the
ratio
\begin{displaymath}
{{d\sigma \over dt }(\gamma \gamma \to \pi^0 \pi^0) / {d\sigma
\over dt}(\gamma \gamma \to \pi^+ \pi^-)}
\end{displaymath}
is particularly sensitive to the shape of pion distribution
amplitude.  At fixed pair mass, and high photon virtuality, one can
study the distribution amplitude of multi-hadron states
\cite{Diehl:2000uv}.  Two-photon annihilation will provide much
information on fundamental QCD processes such as deeply virtual
Compton scattering and large angle Compton scattering in the
crossed channel.  I have also emphasized the interrelation between
the wavefunctions measured in two-photon collisions and the
wavefunctions needed to study exclusive $B$ and $D$ decays.

Much of the most interesting two-photon annihilation physics is
accessible at low energy,  high luminosity $e^+ e^-$ colliders,
including measurements of channels important in the light-by-light
contribution to the muon $g$--2 and the study of the transition
between threshold production controlled by low-energy effective
chiral theories and the domain where leading-twist perturbative
QCD becomes applicable.

The threshold regime of hadron production in photon-photon and
$e^+ e^-$ annihilation, where hadrons are formed at small relative
velocity, is particularly interesting as a test of low energy
theorems, soliton models, and new types of resonance production.
Such studies will be particularly valuable in double-tagged
reactions where polarization correlations, as well as the photon
virtuality dependence, can be studied.

\section{New Calculational Methods}

The light-front quantization of gauge theories can be carried out in an
elegant way using the Dirac method to impose the light-cone gauge
constraint and eliminate dependent degrees of
freedom~\cite{Srivastava:2000cf}.  Unlike the case in equal-time
quantization, the vacuum remains trivial.  Since only physical degrees of
freedom appear, unitarity is maintained.  One
can verify the QCD Ward identities for the physical light-cone gauge and
compute the QCD
$\beta$ function.  Recently, Srivastava and I~\cite{Srivastava:2002mw} have
extended the light-front quantization procedure to the Standard Model.
The spontaneous symmetry breaking of the gauge symmetry is due to a zero
mode of the scalar field rather than vacuum breaking.  The Goldstone
component of the scalar field provides mass to the $W^\pm$ and
$Z^0$ gauge bosons as well as completing its longitudinal
polarization.  The resulting theory is free of Faddeev-Popov ghosts
and is unitary and renormalizable.  The resulting rules give an elegant
new way to compute Standard model processes using light-front Hamiltonian
theory.

The light-front method thus suggests the possibility of developing
an ``event amplitude generator" for high energy processes such as
photon-photon collisions by calculating amplitudes for specific parton
spins using light-front time-ordered perturbation
theory~\cite{Brodsky:2001ww}.  The positivity of the $k^+$ light-front
momenta greatly constrains the number of contributing light-front time
orderings.  The renormalized amplitude can be obtained diagram by diagram
by using the ``alternating denominator" method
which automatically subtracts the relevant counterterm.  The DLCQ method
also provides a simple way to discretized the light-front momentum
variables, while maintaining frame-independence.  The resulting
renormalized amplitude can be convoluted with the light-front
wavefunctions to simulate hadronization and hadron matrix elements.

\section*{Acknowledgments}
I thank Professor Clem Heusch for organizing and inviting me to
this workshop and to David Asner, Jack Gunion, Dae Sung Hwang, Tom
Rizzo, Ivan Schmidt, and Peter Zerwas for helpful comments. Work
supported by the Department of Energy under contract number
DE-AC03-76SF00515.


\begin{thebibliography}{99}

\bibitem{Brodsky:1994nf}
For a review see, S.~J.~Brodsky and P.~M.~Zerwas,
Nucl.\ Instrum.\ Meth.\ A {\bf 355}, 19 (1995)
[arXiv:hep-ph/9407362].


\bibitem{Ginzburg:1981ik}
I.~F.~Ginzburg, G.~L.~Kotkin, V.~G.~Serbo and V.~I.~Telnov,
JETP Lett.\  {\bf 34},
491 (1981) [Pisma Zh.\ Eksp.\ Teor.\ Fiz.\  {\bf 34}, 514 (1981)].

\bibitem{Telnov:1998vs}
V.~Telnov,
{\it 17th International Conference on High-Energy Accelerators
(HEACC 98), Dubna, Russia, 7-12 Sep 1998} [arXiv:hep-ex/9810019].

\bibitem{Asner:2001vu}
D.~Asner, S.~Boege, J.~Early, J.~Gronberg, K.~Skulina, K.~van
Bibber and T.~Markiewicz,
PAC-2001-FPAH055 {\it Presented at IEEE Particle Accelerator
Conference (PAC2001), Chicago, Illinois, 18-22 Jun 2001}.

\bibitem{Velasco:2002vg}
M.~M.~Velasco {\it et al.},
in {\it Proc. of the APS/DPF/DPB Summer Study on the Future of
Particle Physics (Snowmass 2001) } ed. R.~Davidson and C.~Quigg,
arXiv:hep-ex/0111055.

\bibitem{Asner:2001ia}
See, e.g.,  D.~M.~Asner, J.~B.~Gronberg and J.~F.~Gunion,
[arXiv:hep-ph/0110320].

\bibitem{Grzadkowski:1992sa}
B.~Grzadkowski and J.~F.~Gunion,
Phys.\ Lett.\ B {\bf 294}, 361 (1992) [arXiv:hep-ph/9206262].

\bibitem{Cheung:bn}
K.~Cheung,
[arXiv:hep-ph/9310340].

\bibitem{tang}
Wai-Keung Tang, unpublished.

\bibitem{Boos:2000ki}
E.~Boos {\it et al.},
Nucl.\ Instrum.\ Meth.\ A {\bf 472}, 100 (2001)
[arXiv:hep-ph/0103090].

\bibitem{HwangSjb}
D. S. Hwang and S. J. Brodsky (unpublished).

\bibitem{Asner:2001vh}
D.~Asner {\it et al.},
arXiv:hep-ex/0111056.

\bibitem{Brodsky:1993xp}
S.~J.~Brodsky,
SLAC-PUB-6314 {\it Presented at the 2nd International Workshop on
Physics and Experiments with Linear e+ e- Colliders, Waikoloa, HI,
26-30 Apr 1993}.

\bibitem{Chanowitz:1994aq}
M.~S.~Chanowitz,
Nucl.\ Instrum.\ Meth.\ A {\bf 355}, 42 (1995)
[arXiv:hep-ph/9407231].

\bibitem{Gunion:1992ce}
J.~F.~Gunion and H.~E.~Haber,
Phys.\ Rev.\ D {\bf 48}, 5109
(1993).

\bibitem{Drell:1966jv}
S.~D.~Drell and A.~C.~Hearn,
Phys.\ Rev.\ Lett.\  {\bf 16}, 908 (1966).

\bibitem{Gerasimov:1965et}
S.~B.~Gerasimov,
Sov.\ J.\ Nucl.\ Phys.\  {\bf 2}, 430 (1966)
[Yad.\ Fiz.\  {\bf 2}, 598 (1966)].

\bibitem{Brodsky:1992px}
S.~J.~Brodsky and J.~R.~Hiller,
Phys.\ Rev.\ D {\bf 46}, 2141 (1992).

\bibitem{Tung:kn}
W.~K.~Tung,
Phys.\ Rev.\  {\bf 176}, 2127 (1968).

\bibitem{Brodsky:1980zm}
S.~J.~Brodsky and S.~D.~Drell,
Phys.\ Rev.\ D {\bf 22}, 2236 (1980).

\bibitem{Yehudai:1991az}
E.~Yehudai,
Phys.\ Rev.\ D {\bf 44}, 3434 (1991).

\bibitem{Layssac:2001ur}
J.~Layssac and F.~M.~Renard,
Phys.\ Rev.\ D {\bf 64}, 053018 (2001) [arXiv:hep-ph/0104205].

\bibitem{Jikia:1993pg}
G.~Jikia and A.~Tkabladze,
Phys.\ Lett.\ B {\bf 332}, 441 (1994) [arXiv:hep-ph/9312274].

\bibitem{Gounaris:2001rk}
G.~J.~Gounaris, P.~I.~Porfyriadis and F.~M.~Renard,
Eur.\ Phys.\ J.\ C {\bf 20}, 659 (2001) [arXiv:hep-ph/0103135].

\bibitem{Jikia:1991hc}
G.~V.~Jikia,
Nucl.\ Phys.\ B {\bf 374}, 83 (1992).

\bibitem{Boos:2001sj}
E.~Boos, M.~Dubinin, A.~Pukhov, M.~Sachwitz and H.~J.~Schreiber,
Eur.\ Phys.\ J.\ C {\bf 21}, 81 (2001) [arXiv:hep-ph/0104279].

\bibitem{Brodsky:1995ga}
S.~J.~Brodsky, T.~G.~Rizzo and I.~Schmidt,
Phys.\ Rev.\ D {\bf 52}, 4929 (1995)
[arXiv:hep-ph/9505441].

\bibitem{Rizzo:1999xj}
T.~G.~Rizzo,
[arXiv:hep-ph/9907395].

\bibitem{ginz}
I.  F.  Ginzburg, G.  L.  Kotkin, S.  L.  Panfil and V.  G. Serbo,
{  Nucl.  Phys.} {\bf B228}, 285  (1983).  See also S.  Y.  Choi
and F.  Schrempp, {  Phys.  Lett.}  {\bf B272}, 149 (1991); M.
Raidal, {  Nucl. Phys.}  {\bf B441}, 49 (1995),
 [arXiv:hep-ph/9411243] .

\bibitem{Brodsky:1978rp}
S.~J.~Brodsky, T.~A.~DeGrand, J.~F.~Gunion and J.~H.~Weis,
Phys.\ Rev.\ D {\bf 19}, 1418 (1979).

\bibitem{Drees:1995ti}
M.~Drees and T.~Han,
Phys.\ Rev.\ Lett.\  {\bf 76}, 3076 (1996) [arXiv:hep-ph/9512361].

\bibitem{Brodsky:1999mz}
S.~J.~Brodsky, J.~Rathsman and C.~Merino,
Phys.\ Lett.\ B {\bf 461}, 114 (1999) [arXiv:hep-ph/9904280].

\bibitem{Brodsky:1997sd}
S.~J.~Brodsky, F.~Hautmann and D.~E.~Soper,
Phys.\ Rev.\ D {\bf 56}, 6957 (1997)
[arXiv:hep-ph/9706427].

\bibitem{Bartels:1996ke}
J.~Bartels, A.~De Roeck and H.~Lotter,
Phys.\ Lett.\ B {\bf 389}, 742 (1996)
[arXiv:hep-ph/9608401].

\bibitem{Lipatov:2000me}
A.~V.~Lipatov and N.~P.~Zotov,
{\it Liverpool 2000, Deep Inelastic Scattering,} pp. 157-158

\bibitem{Fadin:1998py}
V.~S.~Fadin and L.~N.~Lipatov,
Phys.\ Lett.\ B {\bf 429}, 127 (1998) [arXiv:hep-ph/9802290].

\bibitem{Brodsky:1998kn}
S.~J.~Brodsky, V.~S.~Fadin, V.~T.~Kim, L.~N.~Lipatov and
G.~B.~Pivovarov,
JETP Lett.\  {\bf 70}, 155 (1999) [arXiv:hep-ph/9901229].

\bibitem{Brodsky:1971vm}
S.~J.~Brodsky, T.~Kinoshita and H.~Terazawa,
Phys.\ Rev.\ Lett.\  {\bf 27}, 280 (1971).

\bibitem{Walsh:xy}
T.~F.~Walsh,
Phys.\ Lett.\ B {\bf 36}, 121 (1971).

\bibitem{Krawczyk:2000nh}
M.~Krawczyk,
{\it Ambleside 2000, Structure and Interactions of the Photon}
[arXiv:hep-ph/0012179].

\bibitem{Witten:ju}
E.~Witten,
Nucl.\ Phys.\ B {\bf 120}, 189 (1977).

\bibitem{Peterson:1982tt}
C.~Peterson, P.~M.~Zerwas and T.~F.~Walsh,
Nucl.\ Phys.\ B {\bf 229}, 301 (1983).

\bibitem{Brodsky:2001ue}
S.~J.~Brodsky, P.~Hoyer, N.~Marchal, S.~Peigne and F.~Sannino,
[arXiv:hep-ph/0104291].

\bibitem{Brodsky:2002cx}
S.~J.~Brodsky, D.~S.~Hwang and I.~Schmidt,
Phys.\ Lett.\ B {\bf 530}, 99 (2002)
[arXiv:hep-ph/0201296].

\bibitem{Chernyak:xe}
V.~L.~Chernyak and I.~R.~Zhitnitsky,
Nucl.\ Phys.\ B {\bf 222}, 382 (1983).

\bibitem{Ginzburg:gy}
I.~F.~Ginzburg, D.~Y.~Ivanov and V.~G.~Serbo,
Phys.\ Atom.\ Nucl.\  {\bf 56}, 1474
(1993) [Yad.\ Fiz.\  {\bf 56N11}, 45 (1993)].

\bibitem{Ashery:1999nq}
D. Ashery,  [E791 Collaboration],
[hep-ex/9910024].

\bibitem{BrodskyLepage}
S. J. Brodsky  and G. P. Lepage,  {Phys. Rev.  Lett.} {\bf 53},
545 (1979); {Phys. Lett.} {\bf 87B}, 359 (1979); G. P. Lepage and
S. J. Brodsky, {Phys. Rev.} {\bf D22}, 2157 (1980).

\bibitem{BA}
D. Asner and S. J. Brodsky, unpublished.

\bibitem{Pennington:2000ai}
M.~R.~Pennington,
{\it Ambleside 2000, Structure and Interactions of the Photon}
388-393, [arXiv:hep-ph/0009267].

\bibitem{Acciarri:2001ex}
M.~Acciarri  {\em et al.}  [L3 Collaboration],
Phys.\ Lett.\ B {\bf 501}, 173 (2001) [hep-ex/0011037].

\bibitem{Heister:2002if}
A.~Heister {\it et al.}  [ALEPH Collaboration],
Phys.\ Lett.\ B {\bf 530}, 56 (2002) [arXiv:hep-ex/0202011].

\bibitem{Brodsky:1981rp}
S. J. Brodsky  and G. P.  Lepage,
{  Phys.\ Rev.\ } {\bf D24}, 1808 (1981).

\bibitem{Muller:1994fv}
D. Muller, D. Robaschik, B. Geyer, F. M. Dittes, and J. Horejsi,
{  Fortsch.\ Phys.}   {\bf 42}, 101 (1994), [hep-ph/9812448].

\bibitem{Diehl:2000uv}
M.~Diehl, T.~Gousset and B.~Pire,
 Phys.\ Rev.\ D {\bf 62}, 073014 (2000) [hep-ph/0003233].
M.~Diehl, T.~Gousset, B.~Pire and O.~Teryaev,
Phys.\ Rev.\ Lett.\ {\bf 81}, 1782 (1998) [arXiv:hep-ph/9805380].

\bibitem{Brodsky:1971ud}
S.~J.~Brodsky, T.~Kinoshita and H.~Terazawa,
Phys.\ Rev.\ D {\bf 4}, 1532 (1971).

\bibitem{Sommermann:1992yh}
H. Sommermann, M. R.  Seki, S. Larson.   and S. E. Koonin,
{  Phys.\ Rev.\ }  {\bf D45}, 4303 (1992); S. J. Brodsky  and M.
Karliner,(unpublished).

\bibitem{Karliner:2002nk}
M.~Karliner and S.~Nussinov,
arXiv:hep-ph/0202234.


\bibitem{Brodsky:1985cr}
S.~J.~Brodsky and C.~R.~Ji,
Phys.\ Rev.\ Lett.\  {\bf 55}, 2257 (1985).

\bibitem{Hwang}
{D.~S.~Hwang, these proceedings}.


\bibitem{Brodsky:2000xy}
S.~J.~Brodsky, M.~Diehl and D.~S.~Hwang,
Nucl.\ Phys.\ B {\bf 596}, 99 (2001) [arXiv:hep-ph/0009254].


\bibitem{Brodsky:1997dh}
S.~J.~Brodsky, C.~R.~Ji, A.~Pang and D.~G.~Robertson,
Phys.\ Rev.\ D {\bf 57}, 245 (1998) [arXiv:hep-ph/9705221].



\bibitem{Gronberg:1998fj}
J. Gronberg    {\em et al.} [CLEO Collaboration], {  Phys. Rev.}
{\bf D57}, 33 (1998), [hep-ex/9707031].


\bibitem{Dalley:2000dh}
S.~Dalley,
Nucl.\ Phys.\ Proc.\ Suppl.\  {\bf 90}, 227 (2000)
[arXiv:hep-ph/0007081].



\bibitem{Aitala:2001hb}
E.~M.~Aitala  {\em et al.}  [E791 Collaboration],
 {Phys.\ Rev.\ Lett.}
{\bf 86}, 4768 (2001) [arXiv:hep-ex/0010043].

\bibitem{Radyushkin}
A. V. Radyushkin,  {  Acta Phys.  Polon.} {\bf B26}, 2067 (1995).

\bibitem{Ong}
S. Ong,  {  Phys. Rev.} {\bf D52}, 3111 (1995).

\bibitem{Kroll}
P. Kroll   and M. Raulfs,   {  Phys. Lett.}  {\bf 387B}, 848
(1996).

\bibitem{CZ}
V. L. Chernyak  and  A. R. Zhitnitsky,  {  Phys. Rep.}  {\bf 112},
173 (1984).

\bibitem{Khodjamirian:1999tk}
A.~Khodjamirian,
Eur.\ Phys.\ J.\ C {\bf 6}, 477 (1999) [arXiv:hep-ph/9712451].

\bibitem{Schmedding:2000ap}
A.~Schmedding and O.~Yakovlev,
Phys.\ Rev.\ D {\bf 62}, 116002 (2000) [arXiv:hep-ph/9905392].


\bibitem{Bakulev:2001pa}
A.~P.~Bakulev, S.~V.~Mikhailov and N.~G.~Stefanis,
Phys.\ Lett.\ B {\bf 508}, 279 (2001) [arXiv:hep-ph/0103119].



\bibitem{BHS}
A. Szczepaniak, E. M. Henley,  and S. J. Brodsky, {  Phys. Lett.}
{\bf B243}, 287 (1990).

\bibitem{Sz}
A. Szczepaniak,  {  Phys. Rev.} {\bf D54}, 1167 (1996).


\bibitem{Beneke:1999br}
M.~Beneke, G.~Buchalla, M.~Neubert and C.~T.~Sachrajda,
Phys.\ Rev.\ Lett.\  {\bf 83}, 1914 (1999) [arXiv:hep-ph/9905312].




\bibitem{Keum:2000ph}
Y.~Y.~Keum, H.~n.~Li and A.~I.~Sanda,
Phys.\ Lett.\ B {\bf 504}, 6 (2001) [arXiv:hep-ph/0004004].

\bibitem{Keum:2000wi}
Y.~Y.~Keum, H.~N.~Li and A.~I.~Sanda,
Phys.\ Rev.\ D {\bf 63}, 054008 (2001) [arXiv:hep-ph/0004173].




\bibitem{Brodsky:1981kj}
S. J. Brodsky and G. P. Lepage,
 {Phys.\ Rev.}  {\bf D24}, 2848 (1981).

\bibitem{Dominick:1994bw}
J.~Dominick {\em et al.}  [CLEO Collaboration],
Phys.\ Rev.\ D {\bf 50}, 3027 (1994) [hep-ph/9403379];
J. Boyer {\em et al.},
 {Phys.\ Rev.\ Lett.} {\bf 56}, 207 (1980); TPC/Two Gamma
Collaboration (H. Aihara et al),  {Phys. Rev. Lett.} {\bf 57}, 404
(1986).

\bibitem{Diehl:2001fv}
M.~Diehl, P.~Kroll and C.~Vogt,
[arXiv:hep-ph/0112274].

\bibitem{Brodsky:2001jw}
S.~J.~Brodsky,
Published in {\it Ise-Shima 2001, B physics and CP violation}, pp.
229-234. [hep-ph/0104153].

\bibitem{Brodsky:1988xz}
S.~J.~Brodsky and A.~H.~Mueller,
 {Phys.\ Lett.}  {\bf B206}, 685 (1988).


\bibitem{Brodsky:1980ny}
S. J. Brodsky,  Y.  Frishman, G. P.  Lepage, and C. Sachrajda,
 {Phys. Lett.} {\bf 91B}, 239 (1980).

\bibitem{Brodsky:2000cr}
S. J. Brodsky, E.  Gardi, G.  Grunberg,  and J. Rathsman,
{  Phys. Rev.} {\bf D63}, 094017 (2001) [hep-ph/0002065].

\bibitem{Braun:1999te}
V. M. Braun, S. E.  Derkachov, G. P.  Korchemsky,
 and A. N. Manashov,
 {Nucl.\ Phys.}  {\bf B553}, 355 (1999) [arXiv:hep-ph/9902375].

\bibitem{Farrar:1990qj}
G. R. Farrar  and H. Zhang,
 {Phys.\ Rev.\ Lett.}  {\bf 65}, 1721 (1990).

\bibitem{Kronfeld:1991kp}
A. S. Kronfeld and B. Nizic,
 {Phys.\ Rev.}  {\bf D44}, 3445 (1991).

\bibitem{Guichon:1998xv}
P. A. Guichon and M.  Vanderhaeghen,
 {Prog.\ Part.\ Nucl.\ Phys.}  {\bf 41}, 125 (1998),
[hep-ph/9806305].


\bibitem{Brooks:2000nb}
T.~C.~Brooks and L.~J.~Dixon,
Phys.\ Rev.\ D {\bf 62}, 114021 (2000) [arXiv:hep-ph/0004143].



\bibitem{Brodsky:1972vv}
S. J. Brodsky, F. E.  Close, and J. F.  Gunion,  {Phys. Rev.} {\bf
D6}, 177 (1972).



\bibitem{Pobylitsa:2001cz}
P.~V.~Pobylitsa, .~V.~Polyakov and M.~Strikman,
Phys.\ Rev.\ Lett.\  {\bf 87}, 022001 (2001)
[arXiv:hep-ph/0101279].




\bibitem{Burkardt:2001mf}
M.~Burkardt and S.~K.~Seal,
Phys.\ Rev.\ D {\bf 65}, 034501 (2002) [arXiv:hep-ph/0102245].



\bibitem{Srivastava:2000cf}
P.~P.~Srivastava and S.~J.~Brodsky,
Phys.\ Rev.\ D {\bf 64}, 045006 (2001) [arXiv:hep-ph/0011372].

\bibitem{Srivastava:2002mw}
P.~P.~Srivastava and S.~J.~Brodsky,
[arXiv:hep-ph/0202141].

\bibitem{Brodsky:2001ww}
S.~J.~Brodsky,
[arXiv:hep-th/0111241].

\end{thebibliography}
\end{document}